\documentstyle[preprint,aps,prl,epsf,epsfig]{revtex}
\begin{document}


\title{
Weakly nonlinear theory of grain boundary motion in patterns with
crystalline symmetry
}
\author{Denis Boyer}
\address{Instituto de F\'\i sica, Universidad
Nacional Aut\'onoma de M\'exico, Apartado Postal 20-364, 
01000 M\'exico D.F., M\'exico
}
\author{Jorge Vi\~nals}
\address{Laboratory of Computational Genomics, Donald Danforth Plant Science
Center, 975 North Warson Rd, St. Louis, Missouri 63132.}

\date{\today}

\maketitle

\begin{abstract} 
We study the motion of a grain boundary separating two otherwise
stationary domains of hexagonal symmetry.
Starting from an order parameter equation appropriate for hexagonal 
patterns, a multiple scale analysis leads to an analytical equation
of motion for the boundary that shares many properties with that of
a crystalline solid. We find that
defect motion is generically opposed by a pinning force that arises
from non-adiabatic corrections to the standard amplitude equation. The
magnitude of this force depends sharply 
on the mis-orientation angle between adjacent domains so that
the most easily pinned grain boundaries are those with an angle 
$4^{\circ}\le \theta\le 8^{\circ}$. Although
pinning effects may be small, they do not vanish asymptotically
near the onset of this subcritical bifurcation, and can be orders of 
magnitude larger than those present in smectic phases that bifurcate 
supercritically.
\end{abstract}



The microstructure of a condensed phase and the distribution of topological 
defects largely determine its mechanical and thermodynamical response, as well 
as the temporal evolution of its non-equilibrium configurations. 
Continuum or hydrodynamic approaches to phases with broken
symmetry are now well understood \cite{re:martin72,re:chaikin95},
including the long wavelength description of topological defects
\cite{re:kosevich79,re:nelson79}. Nevertheless, a quantitative theory
of defect motion (e.g., dislocation glide) remains a difficult
problem because it requires short scale phenomena 
that lie beyond a hydrodynamic theory, and therefore beyond the degree 
of universality that such a description entails.

Bifurcations to states with broken symmetry 
are also encountered in a variety of physical, chemical and biological systems 
driven outside of thermodynamic equilibrium. 
The characteristic length scales involved are much larger than atomic 
dimensions, therefore making the observation and study of defects substantially 
easier. Here, amplitude or phase equations that focus on slow modulations of 
a base periodic pattern play the role of the hydrodynamic description
\cite{re:cross93}. It has been noted, for example, that at leading order
hexagonal patterns are the dissipative analogues of a two dimensional, 
isotropic solid \cite{re:walgraef96}.
We focus in this paper on grain boundary motion in hexagonal patterns, and 
find many qualitative and quantitative similarities with grain
boundary motion in crystalline solids. In contrast with the latter case,
our starting model allows a detailed analysis of the breakdown of the long 
wavelength description of defect motion through an explicit multiple scale 
analysis. We are then able to derive analytically several results
that are known only qualitatively or empirically in crystalline solids.

Hexagonal patterns formed in a spatially extended system (like in Langmuir
monolayers \cite{re:sagui95} or block copolymer melts \cite{re:elder01})
are often fairly disordered. They consist of randomly 
oriented grains separated by grain boundaries (arrays of 
dislocations in the low angle case), and are equivalent by symmetry to a 
polycrystalline solid. 
This nonequilibrium microstructure usually
evolves on a 
very slow, fluctuation-dependent time scale due to pinning forces
to defect motion. Our focus here is on a coarse grained model of a hexagonal 
pattern, and, in particular, on the motion of a grain boundary separating 
two domains with arbitrary mis-orientation. We extend to this case 
a recent study of defect motion and pinning near a supercritical
bifurcation involving stripe patterns (or smectic phases) \cite{re:boyer02}. 
We show that the coupling between the slow 
variables of an envelope description of the defect and the 
underlying (fast) periodicity of the base pattern leads to
an effective 
periodic potential, the close analogue of the Peierls barrier 
acting on defects in a crystalline solid \cite{re:peierls40}.
Furthermore, we find that the magnitude of the potential barrier does not 
vanish as the subcritical bifurcation point is approached, contrary to
the case of smectic phases \cite{re:boyer02}. 
Hence, not only crystalline phases emerging from a subcritical bifurcation
are harder than smectics in terms of the forces acting on topological defects,
but also an envelope description that ignores the internal degrees of
freedom of defects does not appear to be valid even near the bifurcation 
point. The self-generated pinning effects discussed here are expected to 
be a general feature of modulated phases, and can explain, for example, 
many grain boundary conformations that have been observed experimentally and 
numerically in block copolymer melts \cite{re:netz97}.

We study here the Swift-Hohenberg model of Rayleigh-B\'enard convection
with an additional quadratic term to allow the formation of hexagonal 
patterns \cite{re:cross93,re:walgraef96},
\begin{equation}
\label{sh}
\frac{\partial \psi}{\partial t}=\epsilon \psi-\frac{1}{k_0^4}
(k_0^2+\nabla^2)^2\psi+g_2\psi^2-\psi^3.
\end{equation}
The order parameter $\psi(\vec{x},t)$ is related to the vertical
velocity at the mid plane of the convective cell, $\epsilon$ is the 
reduced Rayleigh number, and $g_{2}$ can be related to deviations
from Boussinesq behavior in the working fluid. The uniform solution
$\psi = 0$ becomes unstable for $\epsilon > 0$ to a periodic pattern of
wavelength $\lambda_{0} = 2\pi/k_{0}$. Hexagonal patterns
are stable for $-|\epsilon_m(g_2)|<\epsilon<\epsilon_M(g_2)$, and roll
patterns for $\epsilon>\epsilon_M$. In this study, we chose $\epsilon
\in [0,\epsilon_M]$ so that only hexagonal patterns are stable.

\begin{figure}
\epsfig{figure=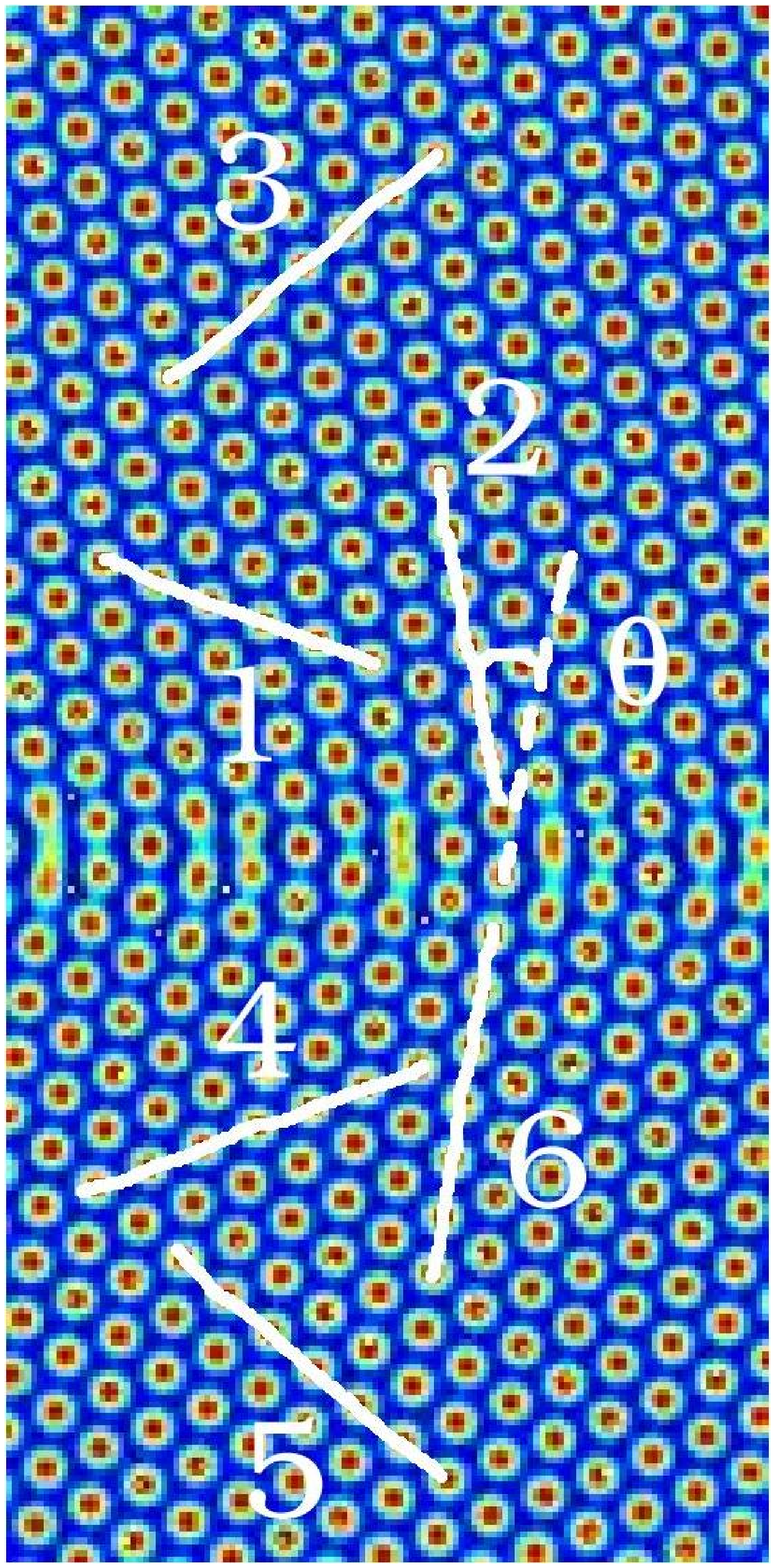,width=2.0in}
\epsfig{figure=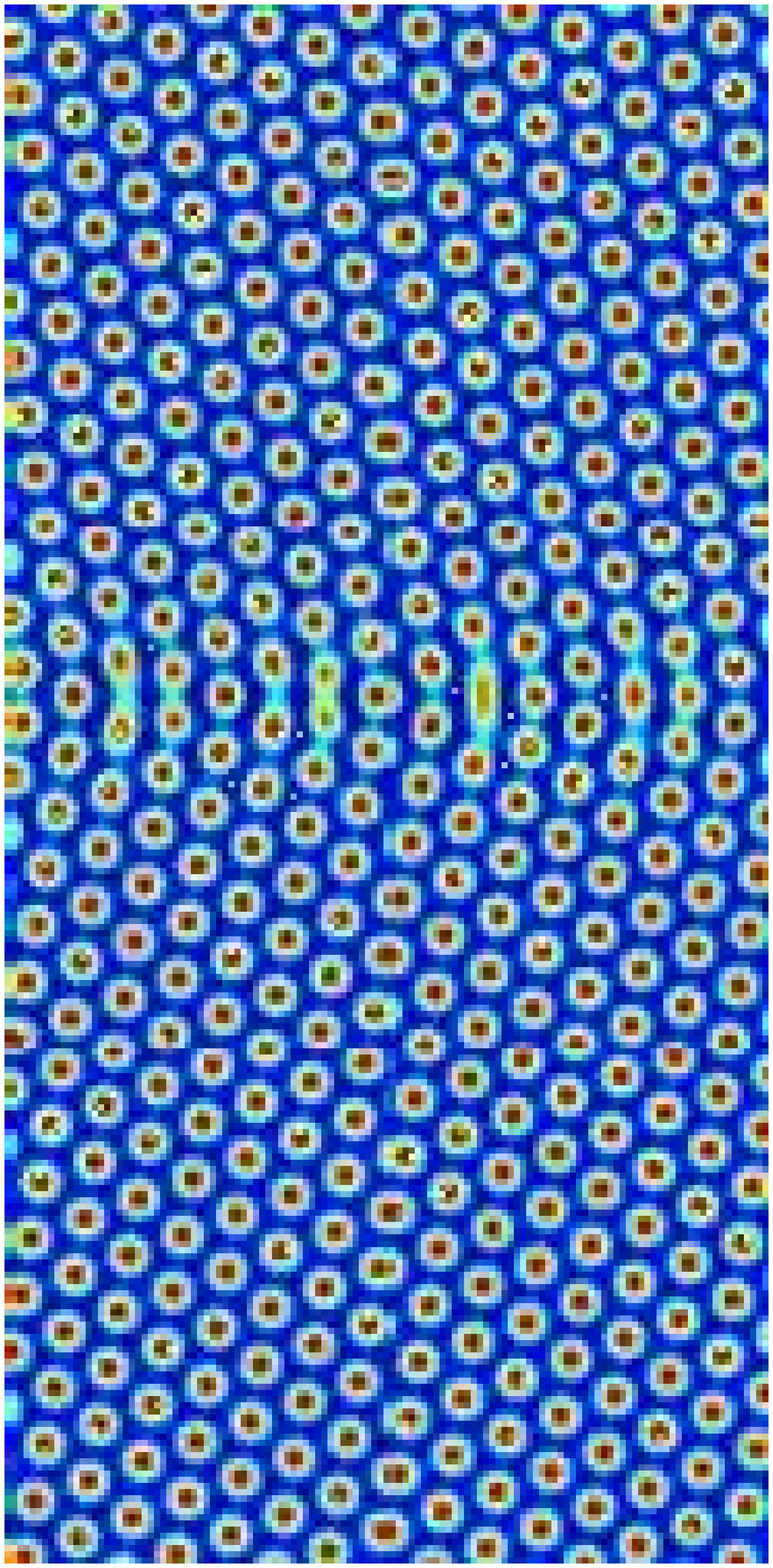,width=2.0in}
\epsfig{figure=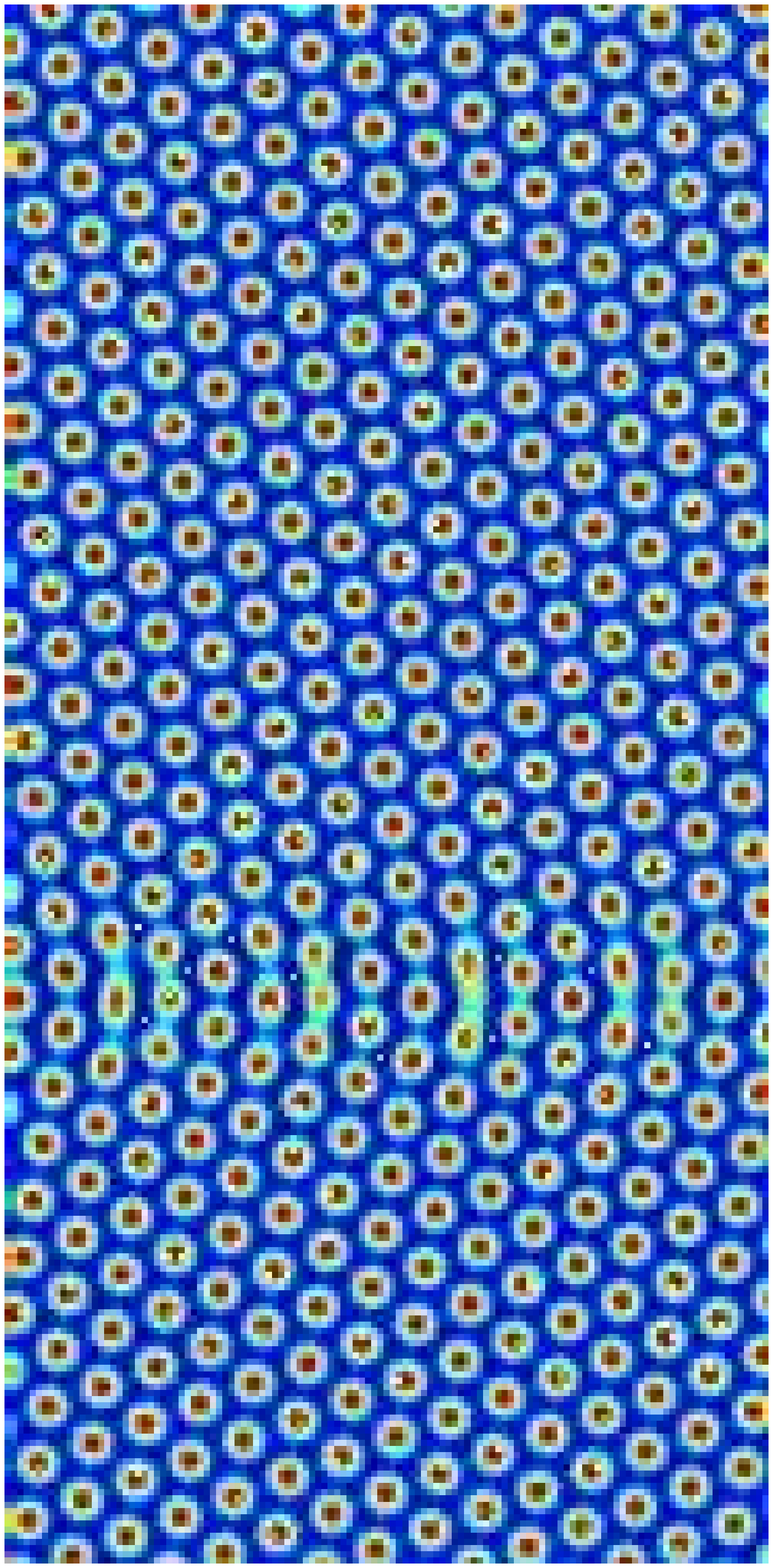,width=2.0in}
\vspace{1.0cm}
\caption{Order parameter $\psi$ in gray scale. Only a portion of a square grid 
of $512^2$ points is shown with spacing $\Delta x=\sqrt{3}/2$
and $\lambda_{0}=8\Delta x$. Three locations are shown in which a planar
grain boundary is stationary.
The two hexagonal domains are defined by the sets
$\{\vec{k}_1,\vec{k}_2,\vec{k}_3\}$ and $\{\vec{k}_4,\vec{k}_5,\vec{k}_6\}$
respectively, with a mis-orientation angle $\theta=\pi/9$.}
\label{figgb}
\end{figure}

An approximate stationary solution for a configuration containing a
planar grain boundary between two uniform and symmetric domains of 
hexagons that have a relative mis-orientation angle $\theta$ 
($0 \le \theta\le\pi/3$, see Fig. \ref{figgb}) can be found by assuming 
that $\psi_0=\sum_{n=1}^{6} A_{n} e^{i\vec{k}_n\cdot \vec{x}}+c.c.$, 
where $A_{n}(x)$ are slowly varying envelopes, and $x$ is 
the coordinate normal to the boundary. 
As $x \rightarrow -\infty$, 
$A_{1,2,3} \rightarrow A_0 =  (g_2+\sqrt{g_2^2+15\epsilon})/15$ 
exponentially fast outside a boundary layer of thickness $\xi$ 
[$A_{n}(x)=f_{n}(x/\xi)$], 
whereas $A_{4,5,6} \rightarrow 0$. As $x \rightarrow \infty$, 
$A_{4,5,6} \rightarrow A_0$ and $A_{1,2,3} \rightarrow 0$. We chose
$|\vec{k}_n|=k_{0}$, $\vec{k}_{1,4}
=k_0(\cos(\phi)\hat{x}\mp\sin(\phi)\hat{y})$, where
$\phi=\pi/6-\theta/2$, and the other vectors $\vec{k}_n$ are obtained 
from these by rotations of $\pm 2\pi/3$ as indicated in Figure \ref{figgb}. 
In analogy with crystalline solids, a small angle grain 
boundary is well described as an array of 
penta-hepta defects (the dislocations cores of a hexagonal pattern), 
separated on 
average by a distance of the order $\lambda_0/\theta$. However, since the 
projections of $\vec{k}_{1,2,3}$ on the $x$ axis are usually not commensurate, 
the patterns observed along the boundary are not periodic. 

We next focus on grain boundary motion and extend our earlier results
for stripe patterns \cite{re:boyer02}. The amplitude equation 
formalism has already been used to study defect dynamics in hexagonal 
patterns \cite{re:tsimring96}. 
Amplitude equations are obtained from a standard multiscale analysis of 
Eq. (\ref{sh}). For example, the equation for $A_1$ follows
from the solvability condition 
\begin{equation}
\label{schex}
\int_x^{x+l_x}\frac{dx'}{l_x}\lim_{l_y\rightarrow\infty}
\int_0^{l_y}\frac{dy'}{l_y}[L(\psi_{0})
+g_2\psi_{0}^2-\psi_{0}^3]e^{-i\vec{k}_1\cdot \vec{x}'}=0,
\end{equation}
where $L$ is a linear operator that follows from Eq. (\ref{sh}) 
\cite{re:manneville90} and $l_x$ a length 
of $O(\lambda_0)$ to be specified later. 
When both $\epsilon$ and $g_2 \rightarrow 0$, the length scale of variation 
of the $A_n$'s ({\it i.e.} the grain boundary thickness $\xi$) 
is much larger than $\lambda_0$, and only non oscillatory terms 
contribute to the integral
(\ref{schex}). The solvability condition leads to the usual amplitude 
equations that show that defects are either immobile or move with constant 
velocity \cite{re:walgraef96}. 

If, on the other hand, $\epsilon$ and $g_2$ are small but finite, 
then the amplitudes are not strictly constant within a period $l_x$.
Any oscillatory term in (\ref{schex}) of wavevector $\vec{K}$ parallel to 
the normal $x$-axis does not integrate to zero, and the
equation for the slowly varying amplitudes cannot be decoupled from the phase 
of the defect (yielding \lq\lq non-adiabatic corrections" 
\cite{re:pomeau86,re:bensimon88b,re:malomed90,re:cross93}).
Although these contributions are small 
(non-analytic in both $\epsilon$ and $g_2$), they may have dramatic effects.
In the case of Eq. (\ref{sh}), non-vanishing terms can arise from the cubic 
nonlinearity, and are proportional to $A_1^2A_4$, 
$\bar{A}_1\bar{A}_4^2$, $A_1A_3A_5$, $A_1\bar{A}_3\bar{A}_5$, $A_1A_2A_6$ and 
$A_1\bar{A}_2\bar{A}_6$. An oscillatory term proportional to $\cos(Kx')$ in 
Eq. (\ref{schex}) contributes to order $\exp(-|K\xi|)$ to the law of
grain boundary of motion (see Eqs. (\ref{phex}) and (\ref{pestim}) below). 
This contribution is simplest
when both $\epsilon \ll 1$ and $g_2 \ll 1$ so that 
$\xi \gg \lambda_0$ and a single mode with lowest $K$ dominates. In
this limit, the slowest non-adiabatic corrections are those proportional to 
$A_1A_2A_6$ and $A_1\bar{A}_2\bar{A}_6$, which have
$K=|\vec{k}_2+\vec{k}_6|=2k_0\sin(\theta/2)$ (see Fig. \ref{figgb}).
The solvability condition (\ref{schex}) now reads
\begin{eqnarray}
\frac{\partial A_1}{\partial t} & = & -\frac{\partial F}{\partial\bar{A}_1} 
\nonumber \\
& - &\int_{x_0}^{x_0+l_x}\frac{dx}{l_x}\ 
6A_1\left(A_2A_6 e^{-2ik_0x\sin(\theta/2)}
+ {\rm c.c.} \right),
\label{A1hex}
\end{eqnarray}
where $l_x=\lambda_0/[2\sin(\theta/2)]$.
The first two terms in Eq. (\ref{A1hex}) represent the standard
amplitude equation with the Lyapunov functional
\begin{eqnarray}
\label{F}
F & = & \int d\vec{r}\left[ -\epsilon\sum_{n=1}^6|A_n|^2
+ \frac{4}{k_0^2} \sum_{n=1}^{6} |D_nA_n|^2 \right.
\nonumber \\
& - & 2g_{2} ( \bar{A}_1\bar{A}_2\bar{A}_3 +
\bar{A}_4\bar{A}_5\bar{A}_6
+A_1A_2A_3+A_4A_5A_6 )
\nonumber \\
& + & \left. \frac{3}{2} \sum_{n=1}^6|A_n|^4
+ 3\sum_{n< m}|A_n|^2|A_m|^2\right],
\end{eqnarray}
with $D_n=\partial/\partial {x_n}$ the derivative along $\vec{k}_n$.
Equations similar to (\ref{A1hex}) can be derived for the remaining amplitudes
$A_{n}$ (not shown).
The last term in the r.h.s. of Eq. (\ref{A1hex}) is new and represents
the dominant non-adiabatic correction in the limits $\epsilon \ll 1$ and
$g_{2} \ll 1$.

In order to derive an equation of motion for the grain boundary
from the equations for the $A_n$, we first
denote by $a_{n}(x)$ ($1\le n\le 6$) the leading order amplitudes of the 
stationary grain boundary, solutions of the one-dimensional coupled 
Ginzburg-Landau equations $\partial F/\partial \bar{a}_n=0$. 
We then look for solutions of the form $A_n(x,t)=a_n(x-x_{gb}(t))$, 
where $x_{gb}(t)$ is the now time-dependent position of the grain boundary
\cite{re:boyer02}. We find,
\begin{equation}
\label{eqmotion}
D\ \dot{x}_{gb}=
-p_{\rm hex}\ \sin[2k_0x_{gb}\sin(\theta/2)],
\end{equation}
where $D=\int_{-\infty}^{\infty} dx\sum_{n=1}^6 (\partial_x a_n)^2$ 
is a friction coefficient, and
\begin{eqnarray}
\label{phex}
p_{\rm hex} & = &{\rm Max}_{\beta} \int_{-\infty}^{\infty}
dx \cos \left[ 2k_0\sin(2\theta)x+\beta \right]
\left\{ 3 \left[ a_2\partial_x(a_2^2a_6) \right. \right.
\nonumber \\
& + & \left. a_6\partial_x(a_6^2a_2)] \right]
+12 \left[ a_1\partial_x(a_1a_2a_6)+a_3\partial_x(a_3a_2a_6) \right.
\nonumber \\
& + & \left. \left. a_4\partial_x(a_4a_2a_6)+a_5\partial_x(a_5a_2a_6) 
\right] \right\}
\end{eqnarray}
is the (dimensionless) amplitude of a pinning force of wavelength
$\lambda_{0}/[2\sin(\theta/2)]$. This periodic force explicitly arises 
from the coupling between the slowly varying wave amplitudes and the 
periodicity of the base state in the integral term of Eq. (\ref{A1hex}).
Equations (\ref{eqmotion}) and (\ref{phex}) show that
a planar grain boundary initially located at an arbitrary position
relaxes toward the nearest minimum of the periodic 
pinning potential.
The stationary and stable positions of the boundary are thus discrete, 
separated from each other by a distance 
$\Delta x_{gb}=\lambda_0/[2\sin (\theta/2)]>\lambda_0$. 
The two wavevectors with the smallest projection on the grain 
boundary normal ($\vec{k}_2$ and $\vec{k}_6$, see Figure \ref{figgb}) 
set the wavevector ($\vec{k}_2+\vec{k}_6$) of the periodic pinning 
potential.
The usual amplitude equation formalism would instead predict that
$p_{\rm hex}=0$, and $x_{gb}$ is arbitrary and decoupled from the phase 
of the pattern. Figure \ref{figgb} shows three successive stable
locations of the the grain boundary obtained by numerically solving Eq.
(\ref{sh}).  The values of $\Delta x_{gb}$ determined numerically 
agree very well with the analytic result. 

We expect equations of the form of Eq. (\ref{eqmotion}) to be generic at
a bifurcation, and not limited to hexagon-hexagon grain boundaries. 
In particular, we anticipate that non-adiabatic effects are important in 
block copolymer melts, and can explain the conformations of
planar interfaces observed in these systems \cite{re:netz97}.

Remarkably, the present result for a pattern of hexagonal symmetry 
is analogous to the Peierls force acting 
on dislocations in crystalline solids \cite{re:peierls40}. Peierls
calculated the energy of a single dislocation by summing over the 
interactions between atoms within two neighboring layers, their displacements 
given by continuous elasticity as a first approximation.
The energy of the dislocation oscillates as a function of its position
so that it can glide only if a force of finite amplitude acts on it
(the Peierls' force). 
Here, we find that a similar force acts on assemblies of dislocations 
organized in arrays. Like in the simpler geometry studied by Peierls,
the amplitude of the pinning force decays exponentially with the spatial 
thickness of the defect $\xi$. From Eq. (\ref{phex}), we find
\begin{equation}
\label{pestim}
p_{\rm hex}\sim c^* A_0^4\ e^{-2k_0\sin(\theta/2)\xi a^*},
\end{equation}
with $c^*$ and $a^*$ dimensionless 
constants of order unity that can {\it a priori} depend on the
mis-orientation $\theta$. At this point, we note an important distinction 
between defect motion in hexagonal patterns that emerge at a subcritical
bifurcation, and in stripe patterns that bifurcate supercritically. The 
supercritical case was discussed in ref. \cite{re:boyer02} for
a $90^{\circ}$ boundary separating two domains of stripes given
by Eq. (\ref{sh}) with $g_2=0$. 
The corresponding pinning force acting on the boundary, $p_{\rm stripe}$, 
satisfies a relation similar to Eq. (\ref{pestim}), with
$\xi \sim 1/\sqrt{\epsilon}$.
Hence $p_{\rm stripe} \sim \exp(-1/\sqrt{\epsilon})\rightarrow 0$, 
as the control parameter $\epsilon\rightarrow 0$.  
In the hexagonal phase, however, $\xi(\epsilon=0)
\simeq 15\lambda_0/(8\sqrt{6}\pi g_2)$ is finite.
Therefore $p_{\rm hex}$, although small, does not vanishes when
$\epsilon \rightarrow 0$ 
(nor when $\epsilon\rightarrow-|\epsilon_m(g_2)|$). 

\begin{figure}
\vspace{0.7cm}
\epsfig{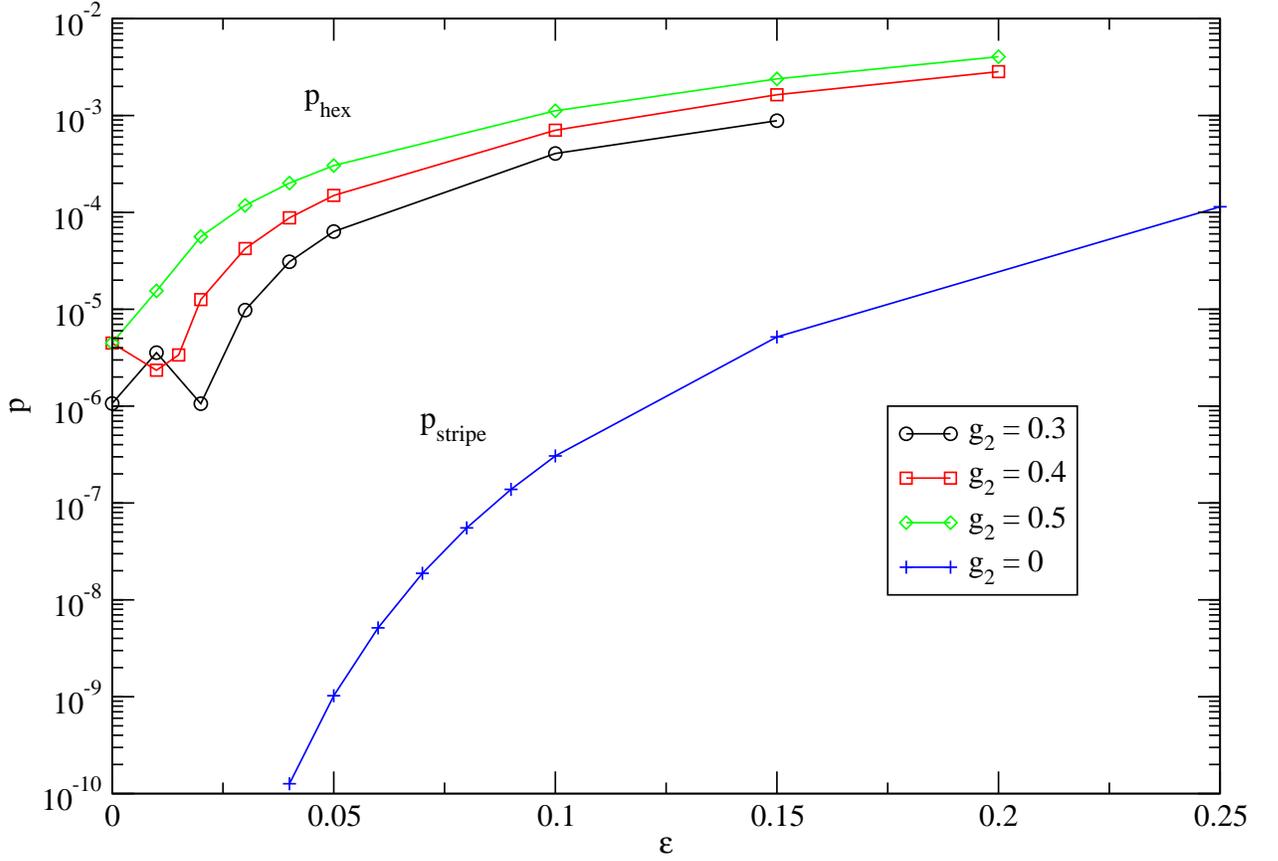}
\vspace{1.0cm}
\caption{Values of the pinning force as given by Eq. (\ref{phex})
as a function of $\epsilon$
for various values of $g_2$ ($\theta=\pi/9$ in each case). The result
corresponding to a stripe pattern was given in ref.
\protect\cite{re:boyer02}, and is shown as a comparison.}
\label{figphex}
\end{figure}

Figure \ref{figphex} shows typical 
variations of the pinning force as a function of $\epsilon$ for different
values of $g_2$ ($\theta=\pi/9$). The curves are given by
Eq. (\ref{phex}), where the amplitudes $a_n$ have been obtained
by numerically integrating the system of 
equations $\partial F/\partial\bar{a}_n=0$.
For a value of $g_2$ as small as $0.3$, $p_{\rm hex}$
is many orders of magnitude larger than $p_{\rm stripe}$. 
Therefore, non-adiabatic effects in hexagonal, \lq\lq crystalline-like", 
patterns 
are difficult to avoid. Defects need to overcome
much higher activation barriers, and will be much less easily un-pinned
either upon the application of external 
stresses or random noise (that would be represented by additional terms 
in the r.h.s. of Eq. (\ref{eqmotion})). 

\begin{figure}
\vspace{0.7cm}
\hspace{-0.3cm}\epsfig{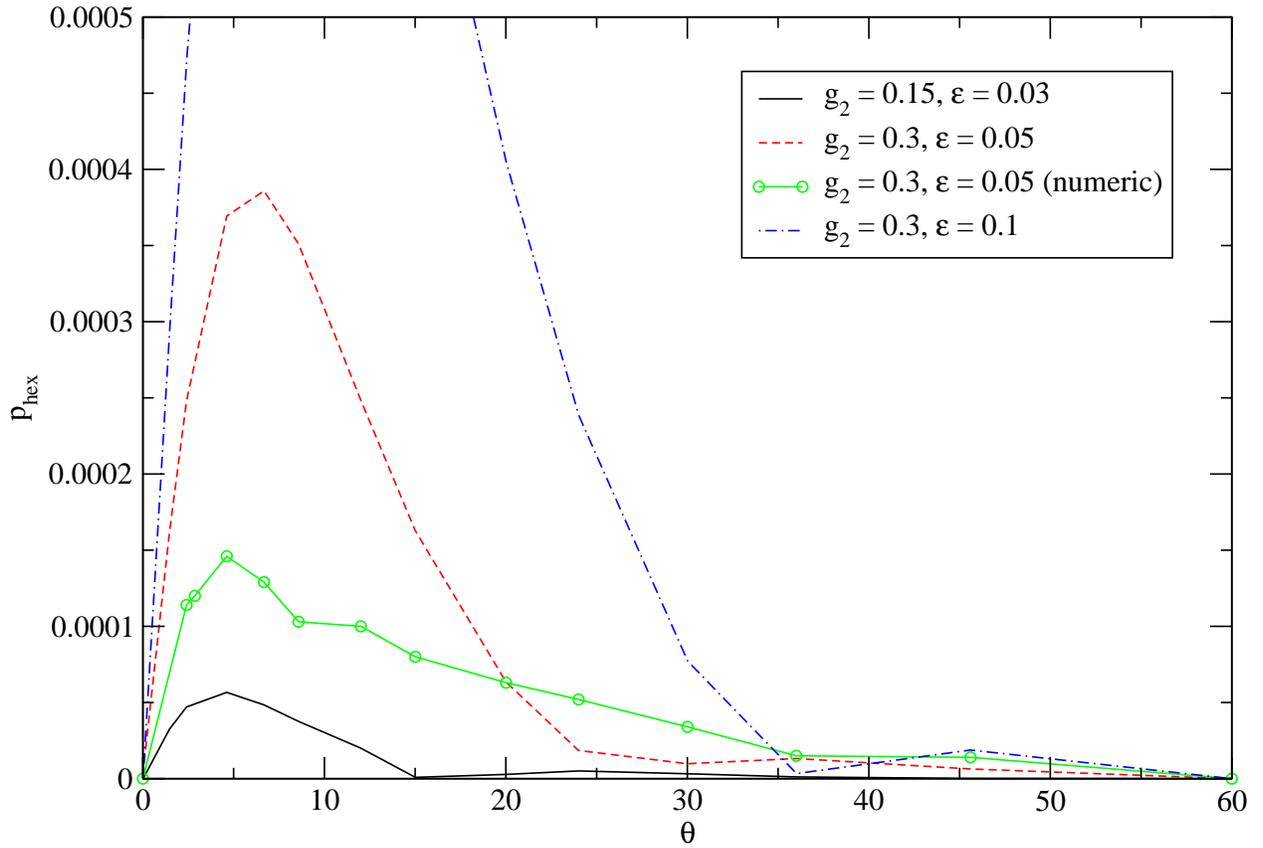}
\vspace{1.0cm}
\caption{Angular dependence of the pinning force for various 
values of $g_2$ and $\epsilon$ obtained analytically with Eq. (\ref{phex}), 
and numerically for one of the parameter sets.}
\label{fi:theta}
\end{figure}

Finally we study the dependence of the pinning force with
grain boundary mis-orientation $\theta$.
Figure \ref{fi:theta} displays $p_{\rm hex}(\theta)$ for different 
values of the parameters $g_2$ and $\epsilon$.
We find that the least mobile grain boundaries, {\it i.e.} those for which 
the amplitude 
of the pinning force is maximal, have a low angle $\theta_M$,
typically such that $4^{\circ} \le \theta_M\le 8^{\circ}$. Both $\theta_M$
and the overall shape of $p_{\rm hex}(\theta)$
depend weakly on $g_2$ and $\epsilon$.
Figure \ref{fi:theta} also shows the pinning force vs. $\theta$ obtained
from a direct numerical solution of Eq. (\ref{sh}) with $g_2=0.3$ and 
$\epsilon=0.05$. Here $p_{\rm hex}$ is estimated by fitting grain boundary
relaxation trajectories to Eq. (\ref{eqmotion}), with the friction 
coefficient $D$ assumed
to be given by the analytic result. The numerical results compare 
reasonably well with the theory given the numerical difficulties in
tracking the relaxation of the grain boundary in a finite sized system.
Although the maximum value is lower and the curve flatter in the numerical 
case, the right order of magnitude is obtained, as well as a similar
range for $\theta_M$. The discrepancy can also be attributed in
part to non-adiabatic corrections of higher order to Eq. (\ref{phex}), that 
may not be negligible at $g_2\sim 0.3$. 

Our present analysis can qualitatively explain the properties of 
polycrystalline, partially ordered configurations that are typically observed 
at long times in many pattern forming systems with this symmetry
\cite{re:sagui95,re:elder01}. 
Although the evolution defined by Eq. (\ref{sh}) is driven by the 
minimization of a Lyapunov functional,
we expect that grain boundaries (and other topological defects) will become 
pinned at long times as the driving force 
for microstructure coarsening decreases. Therefore, the system will
eventually reach metastable,
glassy-like configurations that can only order by slow activated 
processes (presumably logarithmic in time, as already observed in 
\cite{re:elder01} with a random forcing 
term added to Eq. (\ref{sh})). We additionally note that defects observed
in cold metals are essentially low angle grain boundaries in the
range $5^{\circ}$ and $10^{\circ}$ \cite{re:hull92}, a value that
compares very well with the values obtained by the present theory.
  
In summary, we have analyzed the motion of grain boundaries in hexagonal
patterns from an order parameter equation and extended the standard 
Ginzburg-Landau equation for the slowly varying amplitude to 
incorporate non-analytic corrections. 
Like in crystalline phases,
defect motion is opposed by short range forces with periodicity and 
amplitude that strongly depend on the mis-orientation angle between domains. 
Although small, these pinning forces 
can not be neglected asymptotically at long times in a coarsening system, 
even near onset, and are orders of magnitude 
higher than those produced in patterns of smectic symmetry.

We are indebted to Fran\c{c}ois Drolet for useful discussions.
This research has been supported by the U.S. Department of Energy, contract
No. DE-FG05-95ER14566.

\bibliographystyle{prsty}

\begin{thebibliography}{10}

\bibitem{re:martin72}
P. Martin, O. Parodi, and P. Pershan, Phys. Rev. A {\bf 6},  2401  (1972).

\bibitem{re:chaikin95}
P. Chaikin and T. Lubensky, {\em Principles of condensed matter physics}
  (Cambridge University Press, New York, 1995).

\bibitem{re:kosevich79}
A. Kosevich,  in {\em Dislocations in Solids}, edited by F. Nabarro
  (North-Holland, New York, 1979), Vol.~1, p.\ 33.

\bibitem{re:nelson79}
D.R. Nelson and B.I. Halperin, Phys. Rev. B {\bf 19}, 2457 (1979);
A.P. Young, Phys. Rev. B {\bf 19}, 1855 (1979).

\bibitem{re:cross93}
M.C. Cross and P.C. Hohenberg, Rev. Mod. Phys. {\bf 65},  851  (1993).

\bibitem{re:walgraef96}
D. Walgraef, {\em Spatio-Temporal Pattern Formation} (Springer Verlag, New
  York, 1996).

\bibitem{re:sagui95}
C. Sagui and R.C. Desai, Phys. Rev. E {\bf 49},  2225  (1994).

\bibitem{re:elder01}
K.R. Elder, M. Katakowski, M. Haataja, and M. Grant, cond-mat/0107381.

\bibitem{re:boyer02}
D. Boyer and J. {Vi\~nals}, cond-mat/0110254.

\bibitem{re:peierls40}
R. Peierls, Proc. Phys. Soc. London {\bf 52},  34  (1940).

\bibitem{re:netz97}
R.R. Netz, D. Andelman, and M. Schick, Phys. Rev. Lett. {\bf 79}, 
1058 (1997), and references therein.

\bibitem{re:tsimring96}
L. Tsimring, Physica D {\bf 89},  368  (1996).

\bibitem{re:manneville90}
P. Manneville, {\em Dissipative Structures and Weak Turbulence} (Academic, New
  York, 1990).

\bibitem{re:pomeau86}
Y. Pomeau, Physica D {\bf 23},  3  (1986).

\bibitem{re:bensimon88b}
D. Bensimon, B.I. Shraiman, and V. Croquette, Phys. Rev. A {\bf 38},  5461
  (1988).

\bibitem{re:malomed90}
B.A. Malomed, A.A. Nepomnyashchy, and M.I. Tribelsky, Phys. Rev. A {\bf 42},  7244
  (1990).

\bibitem{re:hull92}
D. Hull and D.J. Bacon, {\em Introduction to Dislocations} (Pergamon Press,
New York, 1992).

\end{thebibliography}

\newpage

\newpage

\newpage

\end{document}